\documentstyle[preprint,aps,eqsecnum,prabib]{revtex}
\tighten
\begin{document}

\input{BoxedEPS.tex}

\SetRokickiEPSFSpecial
\HideDisplacementBoxes

\title{A covariant gauge-invariant three-dimensional
description of relativistic bound-states~\footnote{Dedicated
to Professor J.~A. Tjon on the occasion of his 60th birthday.}}

\author {D.~R. Phillips and S.~J. Wallace
\footnote{Email: phillips@quark.umd.edu, wallace@quark.umd.edu.}}

\address{Department of Physics and Center for Theoretical Physics,\\
University of Maryland, College Park, MD, 20742-4111}

\date{\today}
\maketitle

\begin{abstract}
A formalism is presented which allows covariant three-dimensional
bound-state equations to be derived systematically from
four-dimensional ones without the use of delta-functions. The
amplitude for the interaction of a bound state described by these
equations with an electromagnetic probe is constructed. This amplitude is
shown to be gauge invariant if the formalism is truncated at the same
coupling-constant order in both the interaction kernel of the integral
equation and the electromagnetic current operator.
\end{abstract}

\section {Introduction}

The work of Fleischer and Tjon\cite{FT75,FT77,FT80} over twenty years ago
provided a foundation for use of the Bethe-Salpeter (BS) formalism for
nucleon-nucleon ($NN$) scattering. Calculations of electron-deuteron
scattering using these BS wave functions were then performed by
Zuilhof and Tjon\cite{ZT80,ZT81}. Later work of van Faassen and
Tjon\cite{vFT83,vFT84} developed a one-boson-exchange model for the
coupled $NN-N \Delta$ system based on the BS formalism that provides a
reasonable description of the experimental $NN$ phase shift data up to
about 800 MeV. 

Given this progress in four-dimensional calculations, one might ask
whether three-dimensional calculations of the two-nucleon system are
still useful. Reductions of the BS equation to a three-dimensional
equation are motivated by several considerations.  Firstly,
three-dimensional calculations are simpler and hence allow the
application of models for nucleon-nucleon interactions in three-body
problems, e.g. the three-nucleon system and pion-deuteron scattering.
In contrast, the only work using four-dimensional equations in
three-hadron systems is an analysis for the three-nucleon system, by
Rupp and Tjon\cite{RT88,RT92}, which used separable interactions and a
nonrelativistic treatment of spin degrees of freedom. An analysis of
the three-nucleon system with the same ingredients as the BS analyses
of the two-nucleon problem has not been realized due to the technical
complexities involved.  Secondly, the meson-nucleon dynamics in $NN$
interaction models involves `effective' degrees of freedom that are
designed for use in low-order truncations of the full BS kernel. If
the degrees of freedom are truly ``effective'' then one cannot really
regard the meson-nucleon coupling constants as fundamental (with the
exception of the $\pi NN$ coupling constant).  Given this point of
view it is equally valid to construct an effective nucleon-nucleon
potential in three dimensions, with the understanding that the
coupling constants used will be different to those in the
four-dimensional approach.  Nevertheless, almost equivalent results
for physical observables can be obtained in the three-dimensional and
four-dimensional formalisms by making similar truncations of the scattering kernel
and then adjusting the meson-nucleon coupling constants in both
formalisms so that the empirical $NN$ phase shifts are well described.
Thirdly, while it might appear that relativistic effects cannot be
correctly treated in a three-dimensional approach, in fact all
relativistic effects can, in principle, be studied in the
three-dimensional framework.

If relativistic effects are to be studied in a three-dimensional
formalism some systematic reduction from the four-dimensional Bethe-Salpeter
equation to a three-dimensional integral equation must be employed.
Furthermore, if electromagnetic observables are of interest, this
reduction should be implemented so that gauge invariance is realized
in the three-dimensional approach. In this paper we present a
three-dimensional formalism that is covariant and, after consistent
truncation of the current operator and interaction kernel, is gauge
invariant at any desired order.

Consider the Bethe-Salpeter equation (BSE) for the four-dimensional, 
covariant, two-to-two amplitude $T$~\cite{BS51,GL50,Na50,Sc51A,Sc51B},
\begin{equation}
T=K + K G_0 T,
\label{eq:BSE}
\end{equation}
where $G_0$ is the free two-particle propagator, which in our convention is
\begin{eqnarray}
G_0(p_1',p_2';p_1,p_2) &=& i (2 \pi)^8 \delta^{(4)}(p_1' - p_1)
\delta^{(4)}(p_2' - p_2) G_0(p;P),\\
G_0(p;P)&=&d_1(p_1) d_2(p_2),
\end{eqnarray}
with
\begin{equation}
d_i(p_i)=\frac{\Lambda_i^+({\bf p}_i)}{p^0_i - \epsilon_i ({\bf p}_i) + i \eta}
- \frac{\Lambda_i^-({\bf p}_i)}{p^0_i + \epsilon_i ({\bf p}_i) - i \eta},
\label{eq:di}
\end{equation}
where
\begin{eqnarray}
\epsilon_i({\bf p}_i)&=&\sqrt{{\bf p}_i^2 + m_i^2},\\
\Lambda^{\pm}_i({\bf p}_i)&=&\left \{ 
\begin{array}{ll}
\frac{1}{2 \epsilon_i({\bf p}_i)}, & \mbox{for spin-zero particles,}\\
\frac{\pm \epsilon_i ({\bf p}_i) \gamma^0 - {\bf \gamma}_i\cdot {\bf p}_i+m_i}
{2 \epsilon_i({\bf p}_i)}, & \mbox{for spin-half particles.}
\end{array}
\right.
\label{eq:Lambda}
\end{eqnarray}
Here we have used the standard
center-of-mass and relative four-momenta:
\begin{equation}
p_1=\nu_1 P + p; \qquad \qquad p_2=\nu_2 P - p,
\end{equation}
with $\nu_1 + \nu_2=1$. Throughout this paper, when functions are
written with $P$ and $p$ as their arguments they do not contain
four-momentum conserving delta functions. In comparison, when written
with $p_1$ and $p_2$ as their arguments they do contain such delta
functions. In Eq.~(\ref{eq:di}) $\eta$ is a positive infinitesimal. In
Eq.~(\ref{eq:BSE}), $K$ is the Bethe-Salpeter kernel. In principle, $K$
should include all two-particle irreducible two-to-two Feynman graphs.
Solution of (\ref{eq:BSE}) with the full two-particle irreducible
kernel is impractical and usually the kernel is is truncated to lowest
order in the coupling constant, using the ladder approximation, $K=V$.

A simple way to obtain an approximate three-dimensional equation from
Eq.~(\ref{eq:BSE}) is to ignore the relative-energy dependence of the
interaction $V$ in the loop integral of the ladder BSE.  This leads
directly to the Salpeter equation~\cite{Sa52}, which, if only
positive-energy states are considered, is equivalent to the equation of
Blankenbecler-Sugar~\cite{BbS66} and Logunov-Tavkhelidze~\cite{LT63}
\begin{equation}
T=V_{\rm inst} + V_{\rm inst} \langle G_0 \rangle T,
\label{eq:Salpeter}
\end{equation}
with the three-dimensional propagator
\begin{equation}
\langle G_0 \rangle=\int \frac{dp_0}{2 \pi} G_0(p;P).
\end{equation}
Tjon and collaborators have analyzed various hadronic systems,
including elastic and inelastic electron-deuteron
scattering~\cite{HT89,HT90,HT94}, proton-proton
bremsstrahlung~\cite{Sc97}, and the quark structure of
hadrons~\cite{TT92,TT93,TT94}, using a version of this equation with a
modified three-dimensional propagator. These calculations have
provided much insight into the role of relativity in few-body systems.

In order to avoid the instant approximation in the three-dimensional formalism, it is
necessary to correct for the difference between the full
four-dimensional interaction $V$ (or $K$) and the instant version
thereof. Moreover, Eq.~(\ref{eq:Salpeter}) is usually obtained in the
center-of-mass frame of the two-body system and the manner in which
the amplitude is to be boosted to a frame other than the c.m. frame
must be defined.  Much work to address these issues has been based
upon the quasi-potential formalism in which a delta function
constrains the relative energy of the interacting particles. Here, we
discuss a different reduction to three dimensions that does not possess the
unphysical singularities which arise in quasipotential reductions.

A systematic formalism is found by splitting $K$ into two
pieces, one of which, $K_1$, does not depend on the zeroth component
of relative four-momentum, and the other of which, $K -K_1$,
does. Straightforward manipulations then lead to the following set of
equations (see \cite{PW96} for details):
\begin{eqnarray}
\Gamma &=& \left[ 1 -  (K-K_1) G_0 \right]^{-1} \Gamma _1;\\
\Gamma _1 &=& K_1 \langle {\cal G} \rangle \Gamma _1; \label{eq:RTPT}\\
{\cal G} &=&  G_0 +  G_0 (K-K_1) {\cal G}; \label{eq:GG}\\
\langle {\cal G} \rangle &\equiv& \int \frac {dp_0' \, dp_0}{(2 \pi)^2} 
\, {\cal G}(p_\mu',p_\mu;P).
\label{eq:REint}
\end{eqnarray}
There is now a general way to account for all the relative-energy
integrations that reside in $\langle {\cal G} \rangle$.
Demanding that $K_1$ be chosen such that $K_1 \langle {\cal G} \rangle
=K_1 \langle G_0 \rangle $ produces
\begin{eqnarray}
\Gamma _1 &=& K_1 \langle G_0 \rangle \Gamma _1; \label{eq:3Deqn} \\
K_1 &\equiv& \langle G_0 \rangle ^{-1}  \langle G_0 K {\cal G}
\rangle \langle G_0 \rangle ^{-1}.
\label{eq:K1}  
\end{eqnarray}
This choice of $K_1$ is closely related to that used in the work of
Klein\cite{Kl53,Kl54,KM58,KL74} on three-dimensional reductions of
four-dimensional equations, and to standard time-ordered perturbation
theory.

This formalism was developed in a non-covariant way in
Ref.~\cite{PW96}.  In this paper we first show, in
Section~\ref{sec-Section2}, how a simple modification of this argument
allows one to obtain a covariant formalism. (Note that in contrast to
Ref.~\cite{PW96} we do not discuss the approximate inclusion of
crossed-ladder terms in the kernel of the integral equation here,
although it is straightforward to apply the discussion of
Section~\ref{sec-Section2} to that aspect of Ref.~\cite{PW96} also.)
In Section~\ref{sec-Section3} we take up the main issue of the paper:
construction of a gauge-invariant amplitude in this formalism.  We
review the analysis of gauge invariance for the Green's function of
the Bethe-Salpeter equation:
\begin{equation}
G=G_0 + G_0 K G.
\end{equation}
Gauging such four-dimensional Green's functions is relatively
straightforward. The critical issue is whether gauge invariance can be
maintained within a consistent reduction scheme to three dimensions.
We show how this is done in our approach to three-dimensional
reduction, thereby demonstrating that gauge invariance can be
maintained order-by-order by consistently truncating the kernel of the
bound-state equation and the bound-state current matrix element.

\section{Covariant bound-state equations in three dimensions }

\label{sec-Section2}

In this section we generalize the procedure presented in
Ref.~\cite{PW96} to yield a covariant procedure for
obtaining three-dimensional bound-state equations. While the technique
is essentially that discussed in Ref.~\cite{PW96} we display it here
for completeness and clarity.

Consider the Bethe-Salpeter equation for the bound-state vertex function, 
which we write as:
\begin{equation}
\Gamma(p;P)=\int \frac{d^4p'}{(2 \pi)^4} K(p,p';P) G_0(p';P) \Gamma(p';P).
\label{eq:BSEbd}
\end{equation}
As discussed in the Introduction we now split the kernel $K$ via
\begin{equation}
K(p,p';P)=K_1(p,p';P) + K_2(p,p';P).
\label{eq:Kdecomp}
\end{equation}
However, in contrast to the case discussed in the Introduction, here 
we make this decomposition in a covariant way. To do this we must write
the relative four-momenta $p$ and $p'$ as
\begin{equation}
p=p_{\parallel D} \hat{D} + p_{\perp D}; 
\qquad \qquad 
p'=p'_{\parallel D} \hat{D} + p'_{\perp D}.
\end{equation}
Here, for any four-vector $k$, $k_\parallel=k \cdot \hat{D}$,
with $\hat{D}$ a unit four-vector in the direction of
$D$. The piece of the kernel $K_1$ is then chosen so that it
does not depend on $p_{\parallel D}$ or $p'_{\parallel
D}$. 

The choice of vector $D$ is completely arbitrary and nothing depends
upon it if the kernel used in the three-dimensional equation is treated
exactly. Ultimately, of course, the kernel must be truncated, and
this introduces some dependence on the vector $D$ into the three-dimensional
formalism. This will be discussed further at the end of this section.

If the non-covariant vector
\begin{equation}
\hat{D}=(1,0,0,0),
\label{eq:prev}
\end{equation}
is chosen, then the reduction to three dimensions is exactly that developed in
Ref.~\cite{PW96}.  In order to obtain a covariant formalism for
the bound-state problem, $D = P$ must be chosen, because
$P$ is the only available four-vector of the problem that is conserved.
 With this choice, Eq.~(\ref{eq:Kdecomp}) becomes a
manifestly covariant decomposition of the kernel $K$, since then it is
expressed solely in terms of Lorentz-covariant objects. Obviously,
this choice coincides with Eq.~(\ref{eq:prev}) in the center-of-mass
frame of the two-body system. But, in all other frames, this decomposition
differs from that of Ref.~\cite{PW96}, since it boosts from the c.m.
frame in a covariant way.  We keep $D$ distinct from $P$ in what
follows in order that other choices of $D$ may be considered for 
current matrix elements. 

If $\Gamma_1$ is now defined by
\begin{equation}
\Gamma_1(p;P)=\int \frac{d^4p'}{(2 \pi)^4} K_1(p,p';P) G_0(p';P) \Gamma(p';P),
\end{equation}
it follows that, since $K_1$ does not depend on $p_{\parallel
D}$, neither does $\Gamma_1$. It is then a simple matter to
show that $\Gamma_1$ obeys the equation:
\begin{equation} 
\Gamma_1(p_{\perp D};P)=\int \frac{d^4p' d^4p''}{(2 \pi)^8} 
K_1(p_{\perp D},p_{\perp D}';P) {\cal G} (p',p'';P) \Gamma_1(p_{\perp D}'';P),
\label{eq:hold}
\end{equation}
where the four-dimensional Green's function ${\cal G}$ is defined by the 
integral equation
\begin{equation}
{\cal G}(p',p'';P)=G_0(p';P) \delta^{(4)}(p'-p'') + 
\int \frac{d^4p'''}{(2 \pi)^4} G_0(p';P) K_2(p',p''';P) {\cal G}(p''',p'';P).
\label{eq:calG}
\end{equation}
Since ${\cal G}$ is the only four-dimensional object appearing
on the right-hand side of Eq.~(\ref{eq:hold}), the equation reduces to one
involving only three-dimensional integrals:
\begin{equation}
\Gamma_1(p_{\perp D};P)=\int \frac{d^3p_{\perp D}' 
d^3p_{\perp D}''}{(2 \pi)^6} 
K_1(p_{\perp D},p_{\perp D}';P) \langle {\cal G}
\rangle(p_{\perp D}',p_{\perp D}'';P)
\Gamma_1(p_{\perp D}'';P).
\label{eq:3DIE}
\end{equation}
Here, and in the rest of the paper, the notation $\langle {\cal G}
\rangle$ denotes integration of the Green's function ${\cal G}$ over
four-momentum components parallel to $D$, in order to yield a
three-dimensional quantity:
\begin{equation}
\langle {\cal G} \rangle (p_{\perp D}',p_{\perp D}'';P)\equiv
\int \frac{dp_{\parallel D}' dp_{\parallel D}''}{(2 \pi)^2}
{\cal G}(p',p'';P).
\end{equation}

In order to obtain $\Gamma$ from $\Gamma_1$ the equation
\begin{equation}
\Gamma(p;P)=\Gamma_1(p_{\perp D};P) + \int \frac{d^4p'}{(2 \pi)^4}
K_2(p,p';P) G_0(p';P) \Gamma(p';P).
\label{eq:Gamma}
\end{equation}
must be used. The $\Gamma$ obtained from solving Eq.~(\ref{eq:3DIE})
and using Eq.~(\ref{eq:Gamma}) is identically equal to that found by
solving the original Bethe-Salpeter equation (\ref{eq:BSEbd}).  The
difficulty is that the integral equations (\ref{eq:calG}) and
(\ref{eq:Gamma}) are themselves four-dimensional. So, at this stage we
have not simplified the problem at all.

We now consider whether there is a choice for $K_1$ which simplifies
the form of $\cal G$. In particular, we seek a $K_1$ such that the
propagator $\langle \cal{G} \rangle$ in Eq.~(\ref{eq:3DIE}) may be
replaced by the free propagator $\langle G_0 \rangle$.  In other
words, the contribution of the last term of Eq.~(\ref{eq:calG}) must
vanish once the integrals over $p_{\parallel D}'$ and $p_{\parallel
D}''$ are performed. Writing $K_2=K-K_1$ then leads to the defining
condition:
\begin{equation}
\langle G_0 \rangle (p_{\perp D};P)
K_1(p_{\perp D},p_{\perp D}';P)
\langle G_0 \rangle (p_{\perp D}';P)=
\langle \int \frac{d^4p''}{(2 \pi)^4} G_0(p;P) K(p,p'';P) {\cal G}(p'',p';P)
\rangle \label{eq:K1defn}
\end{equation}
with ${\cal G}$ still defined in terms of $K_1$ by
Eq.~(\ref{eq:calG}).  With this choice of $K_1$ Eq.~(\ref{eq:3DIE})
involves a simple propagator:
\begin{equation}
\Gamma_1(p_{\perp D};P)=\int \frac{d^3p_{\perp D}'}{(2 \pi)^3} 
K_1(p_{\perp D},p_{\perp D}';P) \langle 
G_0 \rangle (p_{\perp D}';P)
\Gamma_1(p_{\perp D}';P),
\label{eq:K1eqn}
\end{equation}
and all the complexity is transferred to the interaction kernel $K_1$. 
We write the equation (\ref{eq:K1eqn}) formally as:
\begin{equation}
\Gamma_1=K_1 \langle G_0 \rangle \Gamma_1. 
\end{equation}

In order to calculate the kernel $K_1$ defined by Eq.~(\ref{eq:K1defn}),
a perturbative expansion of the four-dimensional kernel, $K$, must be made:
\begin{equation}
K=\sum_{i=1}^\infty K^{(2i)}.
\end{equation}
Then, to second order in the coupling constant we see that $K_1$ is given by
\begin{equation}
K_1^{(2)}=\langle G_0 \rangle^{-1} \langle G_0 K^{(2)} G_0 \rangle 
\langle G_0 \rangle^{-1},
\label{eq:K1Klein}
\end{equation}
while at fourth order in the coupling constant we have
\begin{equation}
K_1^{(4)}=\langle G_0 \rangle^{-1} \biggl( \langle G_0 K^{(4)} G_0 \rangle + 
 \langle G_0 K^{(2)} G_0 K^{(2)} G_0 \rangle \biggr) \langle G_0 \rangle^{-1}
- K_1^{(2)} \langle G_0 \rangle K_1^{(2)}.
\label{eq:K14}
\end{equation}

Here the propagator $\langle G_0 \rangle$ is
\begin{equation}
\langle G_0 \rangle (p_{\perp D};P^2)=
\frac{\Lambda_1^+ \Lambda_2^+}{P_{\parallel D} + i \eta - \epsilon_1 
- \epsilon_2} - \frac{\Lambda_1^- \Lambda_2^-}{P_{\parallel D}
- i\eta + \epsilon_1 + \epsilon_2},
\label{eq:ETG0}
\end{equation}
where $\Lambda_i^{\pm}=\Lambda_i^{\pm}({p_i}_{\perp D})$.
These operators are defined as in Eq.~(\ref{eq:Lambda}), but with
$\epsilon_i$ now given by $\epsilon_i({p_i}_{\perp D}) \equiv
\sqrt{m_i^2 - {p_i}^2_{\perp D}}$. In other words,
$\epsilon$'s and $\Lambda$'s are now defined as functions of
${p_i}_{\perp D}$, not ${\bf p}_i$. These definitions then
only coincide with the usual definitions of these quantities in the
two-body center of mass frame. 

The form (\ref{eq:ETG0}) reveals an inconsistency in
Eqs.~(\ref{eq:K1Klein}) and (\ref{eq:K14}). In a spin-half theory the
inverse of $\langle G_0
\rangle$ does not exist, since the propagator has no components
in sectors where one particle is in a positive-energy state and the
other is in a negative-energy state. However, if we only take and use
matrix elements of equations such as (\ref{eq:K1Klein}) and
(\ref{eq:K14}) in $++$ and $--$ states the results given above are
entirely correct. This caveat must be borne in mind when examining the
equations presented below.  The difficulty springs from demanding that
$\langle {\cal G} \rangle=\langle G_0 \rangle$ when the entity on the
right-hand side of this equation does not allow propagation in $+-$ or
$-+$ states. This problem can be completely avoided by use of the
modified free propagator of
Refs.~\cite{HT89,HT90,HT94,TT92,TT93,PW96,MW87,Wa88,WM89}, for which
$\langle G_0 \rangle^{-1}$ is uniquely defined in all $\rho$-spin sectors.
The results of this paper may be extended to that case.  However, in
order to focus our discussion on covariance and gauge invariance, we
have chosen not to discuss this extension of the formalism here.

If we consider a spin-zero or spin-half field theory with a Yukawa
interaction, then the interaction $K_1^{(2)}$ takes the form displayed
in Ref.~\cite{PW96}, up to trivial kinematic replacements:
\begin{eqnarray}
&&\langle G_0 \rangle(p_{\perp D}';P)
K_1^{(2)} (p_{\perp D}', p_{\perp D};P)
\langle G_0 \rangle(p_{\perp D};P)=\nonumber\\
&& \quad \frac{g_1 g_2 {\cal M}}{2 \omega} \left[
\left (\frac{{\Lambda_1^+}' {\Lambda_2^+}'}{P_{\parallel D}-\epsilon_1'-\epsilon_2'}  
+ \frac{{\Lambda_1^-}' {\Lambda_2^+}'}{-\epsilon_1' - \epsilon_1 -
\omega} \right)
\frac{1}{P_{\parallel D} - \epsilon_1 - \epsilon_2' - \omega}
\left(\frac{\Lambda_1^+ \Lambda_2^+}{P_{\parallel D}-\epsilon_1-\epsilon_2} + 
\frac{\Lambda_1^+ \Lambda_2^-}{-\epsilon_2 - \epsilon_2' - \omega}\right) 
\right.\nonumber\\
&&\quad +
\left(\frac{{\Lambda_1^+}'{\Lambda_2^+}'}{P_{\parallel D}-\epsilon_1'-\epsilon_2'}
+\frac{{\Lambda_1^+}'{\Lambda_2^-}'}{-\epsilon_2 - \epsilon_2' -
\omega}\right)
\frac{1}{P_{\parallel D} - \epsilon_1' - \epsilon_2 - \omega}
\left (\frac{\Lambda_1^+ \Lambda_2^+}{P_{\parallel D}-\epsilon_1-\epsilon_2}  +
\frac{\Lambda_1^- \Lambda_2^+}{-\epsilon_1' - \epsilon_1 - \omega} \right) 
\nonumber\\
&& \quad + \frac{{\Lambda_1^-}' {\Lambda_2^-}'}{-P_{\parallel D}-\epsilon_1'-\epsilon_2'}
\left(\frac{1}{-\epsilon_1' - \epsilon_1 - \omega} + 
\frac{1}{-\epsilon_2' - \epsilon_2 - \omega}\right)
\frac{\Lambda_1^+ \Lambda_2^+}{P_{\parallel D}-\epsilon_1-\epsilon_2}\nonumber\\
&& \quad +\left(\frac{{\Lambda_1^-}'{\Lambda_2^-}'}{-P_{\parallel D}-\epsilon_1'-\epsilon_2'}
+ \frac{{\Lambda_1^+}' {\Lambda_2^-}'}{-\epsilon_1 -\epsilon_1'-\omega}\right) 
\frac{1}{-P_{\parallel D}-\epsilon_1-\epsilon_2'-\omega}
\left(\frac{\Lambda_1^- \Lambda_2^-}{-P_{\parallel D}-\epsilon_1-\epsilon_2}
+ \frac{\Lambda_1^- \Lambda_2^+}{-\epsilon_2 - \epsilon_2' - \omega}\right)
\nonumber\\
&&\quad +\left(\frac{{\Lambda_1^-}' {\Lambda_2^-}'}{-P_{\parallel D}-\epsilon_1'-\epsilon_2'}
+ \frac{{\Lambda_1^-}' {\Lambda_2^+}'}{-\epsilon_2 -\epsilon_2'-\omega}\right)
\frac{1}{-P_{\parallel D}-\epsilon_1'-\epsilon_2-\omega}
\left(\frac{\Lambda_1^- \Lambda_2^-}{-P_{\parallel D}-\epsilon_1-\epsilon_2}
+ \frac{\Lambda_1^+ \Lambda_2^-}{-\epsilon_1 - \epsilon_1' - \omega}\right)
\nonumber\\
&& \quad \left. +\frac{{\Lambda_1^+}'{\Lambda_2^+}'}{P_{\parallel D}-\epsilon_1'-\epsilon_2'}
\left(\frac{1}{-\epsilon_1' - \epsilon_1 - \omega} +
\frac{1}{-\epsilon_2' - \epsilon_2 - \omega}\right)
\frac{{\Lambda_1^-} {\Lambda_2^-}}{-P_{\parallel D}-\epsilon_1-\epsilon_2}\right],
\label{eq:Kleinpot}
\end{eqnarray}
where $\epsilon_i' \equiv \sqrt{m_i^2 - {p_i'}_{\perp D}^2}$,
$\Lambda^{\pm \prime}_i=\Lambda_i^{\pm} ({p_i'}_{\perp D})$,
$\omega=\omega(p_{\perp D}-p_{\perp D}')$.

The interaction $K_1$ can be systematically improved by including
the next terms in the coupling constant expansion of
Eq.~(\ref{eq:K1defn}). In general, the $K_1$ defined by Eq.~(\ref{eq:K1defn}) 
is the two-particle irreducible interaction, with two-particle
irreducibility defined with respect to the propagator $\langle G_0
\rangle$. 

As in the Klein approach (see Refs.~\cite{Kl53,Kl54,KM58,KL74}) the
three-dimensional Green's function $\langle G_0 + G_0 T G_0 \rangle$,
where $T$ is the Bethe-Salpeter amplitude, may be expanded as a series
of diagrams. From the arguments given so far in this section, it is now
clear that the Klein rules discussed in Refs.~\cite{PW96,Kl53,Kl54}, will
only be modified by the following kinematical replacements:
\begin{eqnarray}
E &\rightarrow& P_{\parallel D};\\
{\bf p}_i &\rightarrow& {p_i}_{\perp D};\\
\epsilon_i({\bf p}_i) &\rightarrow& \epsilon_i({p_i}_{\perp D}) \equiv 
\sqrt{m_i^2 - {p_i}_{\perp D}^2};\\
\omega({\bf p} - {\bf p}') &\rightarrow& \omega(p_{\perp D} - p_{\perp D}')
\equiv \sqrt{\mu^2 - (p_{\perp D} - p_{\perp D}')^2}.
\end{eqnarray}
Any choice for $K_1$ may therefore be expressed as a set of diagrams
and calculated in a straightforward way.  Once again, we observe that
if the non-covariant choice (\ref{eq:prev}) is made for $D$,
then these replacements have no effect.
Since the Klein rules coincide with those of time-ordered perturbation
theory in $++$ states, an interaction generated using a particular set
of diagrams in the formalism discussed here with this choice for
$D$ will agree with the time-ordered perturbation theory
result for the same diagrams, provided only positive-energy states are
considered.

However, unlike either the time-ordered perturbation theory or Klein
interactions, $K_1$ may be defined so that it boosts in a covariant
way. Consider, for instance, the choice $D=P$. The result of the
above procedure is then a covariant three-dimensional interaction for
use in Eq.~(\ref{eq:K1eqn}). In the c.m. frame, where
$\hat{P}=(1,0,0,0)$ this interaction agrees with the Klein
result. After solving for $\Gamma_1$ in the c.m. frame we can simply
make the replacement ${\bf p} \rightarrow p_{\perp P}$ and thereby
obtain a covariant vertex function in any frame. Once this equation is
defined in one frame a kinematic boost defines it in all other frames.

If $D$ is not chosen to be parallel to the total four-momentum of the
two-body system then vertex functions in an arbitrary frame cannot be
obtained from a kinematical boost of the rest-frame vertex function
with the same $D$.  To see this, consider some arbitrary choice for the
four-vector $D$. Suppose that $P$ is chosen to be a four-vector
describing the deuteron at rest and that we wish to calculate
$\Gamma_1 (k_{\perp D};P')$. Under a Lorentz transformation ${\cal L}$
\begin{equation}
\Gamma_1([{\cal L} k]_{D'};{\cal L} P)=S({\cal L})
\Gamma_1(k_{\perp D};P),
\label{eq:covG1Pt}
\end{equation}
where $D'={\cal L} D$, $S({\cal L})=S_1({\cal L}) S_2({\cal L})$, and
$S_i$ is the unitary representation of the boost ${\cal L}$ acting on
the spin-half particle $i$. Consequently, $\Gamma_1(k_{\perp D};P')$
can be related by a Lorentz boost to $\Gamma_1(k_{\perp D_0};P)$,
where ${\cal L}$ is chosen such that $P'={\cal L} P$, and $D_0 \equiv
{\cal L}^{-1} D$. Therefore $\Gamma$ can always be related to {\it a}
rest-frame vertex function, but, unless $D$ is chosen to be the total
four-momentum of the two-body system, this rest-frame vertex
function corresponds to a reduction of the four-dimensional
Bethe-Salpeter vertex different from that used in calculating
$\Gamma_1(k_{\perp D};P)$.

When a truncated kernel $K_1$ is used this difference matters, since
then for each choice of $D$ the kernel of the bound-state equation
(\ref{eq:K1eqn}) is different. Consequently, the bound-state equation
must be re-solved at each value of $P'$ in order to determine the
vertex function $\Gamma(k_{\perp D};P')$. If a truncation of the
kernel $K_1$ is employed, the solutions for values of $P' \neq P$ will
not correspond to the same bound-state mass as the vertex function
$\Gamma_1(k_{\perp D};P)$.  It is possible to minimize this effect by
using a sufficiently high-order truncation of the kernel.  A simpler
expedient is to use a low-order truncation and correct this by
rescaling the kernel as a function of $D$ so as to
maintain the same invariant mass for the initial and final states of
Eq.~(\ref{eq:3damplfull}). Because of Eq.~(\ref{eq:covG1Pt}) this rescaling
may equivalently be viewed as a function of the total four-momentum $P'$.

\section{Construction of a three-dimensional gauge invariant amplitude}

\label{sec-Section3}

In Section~\ref{sec-Section2} we showed how to derive a
three-dimensional two-body bound-state equation from the corresponding
four-dimensional Bethe-Salpeter equation.  With the choice $D=P$ this
method maintains the Lorentz invariance of the original theory.  The
calculations of Refs.~\cite{PW96,LA97} indicate that this formalism
holds promise for accurately approximating the results of
Bethe-Salpeter equations.  However, in any reduction of a
gauge-invariant, four-dimensional theory to three dimensions, a
critical test of the reduction procedure is whether it can
consistently maintain the gauge invariance of the full
four-dimensional theory. In this section we first consider the Green's
function for the interaction of two particles with the photon
constructed from the Bethe-Salpeter Green's function.  The general
construction of currents guarantees that this four-dimensional Green's
function obeys a Ward-Takahashi identity (WTI), as required for gauge
invariance.  (Obtaining such a WTI in the four-dimensional theory in
general requires the inclusion of meson-exchange currents.) We then
turn our attention to the corresponding three-dimensional formalism,
and show that truncating the two-body current and the two-body
interaction $K_1$ at the same order in the coupling constant leads to
a formalism for the calculation of electromagnetic processes which is
both gauge invariant and Lorentz invariant.

We begin our discussion by reviewing the WTI of the Bethe-Salpeter Green's
function.  Consider the two-body Green's function $G$ that 
is the solution of the Bethe-Salpeter equation  
\begin{equation}
G=G_0 + G_0 K G.
\label{eq:GBSE}
\end{equation}
Define the Green's function for the interaction of a free single
particle $i$ with a photon of fixed momentum $Q$ via:
\begin{equation}
g^{(i)}_{0 \mu}(p_i,Q)=d_i(p_i + Q) j^{(i)}_\mu(Q^2) d_i(p_i).
\label{eq:obgamma}
\end{equation}
Here the operator $j_\mu^{(i)}$ is assumed to be such that this
one-body photon Green's function obeys a WTI~\cite{Wa50,Ta57}
\begin{equation}
Q^\mu g^{(i)}_{0 \mu}(p_i,Q)=e_i (d_i(p_i) - d_i(p_i+Q)).
\label{eq:obWTI}
\end{equation}
Defining the two-body analog of
Eq.~(\ref{eq:obgamma}):
\begin{eqnarray}
&&G_0(p_1 + Q,p_2) j^{(1)}_\mu(Q^2)
d_2^{-1}(p_2) G_0(p_1,p_2) + G_0(p_1,p_2+Q) d_1^{-1}(p_1)
j^{(2)}_\mu(Q^2) G_0(p_1,p_2)\nonumber\\
&& \qquad \qquad \equiv G_0(p_1 + Q,p_2) j^{(1)}_\mu(Q^2)
d_2^{-1}(p_2) G_0(p_1,p_2) + (1 \leftrightarrow 2)
\equiv {G_{0}^\gamma}_\mu(p_1,p_2,Q).
\label{eq:tbgamma}
\end{eqnarray}
Here, and throughout the rest of the paper, the notation $(1
\leftrightarrow 2)$ indicates that the momenta of the two
particles must be swapped, {\it and} the labels exchanged. Therefore,
the $(1 \leftrightarrow 2)$ pieces of any expression represent
the photon coupling to whichever particle it did not couple to in
the first part of the expression.  We observe that the
Green's function ${G_0}^\gamma_\mu$ also obeys a WTI
\begin{equation}
Q^\mu {G^\gamma_0}_\mu(p_1,p_2,Q)=e_1 (G_0(p_1,p_2) - G_0(p_1+Q,p_2))
+ (1 \leftrightarrow 2).
\label{eq:G0gamma}
\end{equation}

Now let $G_\mu^\gamma$ be the Green's function for the interaction of
one photon with the interacting two-particle system.  Note that in in
the two-nucleon system the quantities $e_i$ will include isospin
operators, so care must be exercised in ordering charges and any
interactions which also involve isospin operators.  We may write the
following equation for a gauge invariant $G_\mu^\gamma$, by simply
allowing the photon to be inserted anywhere on the right-hand side of
Eq.~(\ref{eq:GBSE}),
\begin{equation}
G^\gamma_\mu={G^\gamma_0}_\mu + {G^\gamma_0}_\mu K G + G_0 K G^\gamma_\mu 
+ G_0 K^\gamma_\mu G.
\label{eq:Ggameq}
\end{equation}
Here $K^\gamma_\mu$ is found by coupling the photon to every internal
charged line in the kernel $K$.

Using Eq.~(\ref{eq:GBSE}) and the definition (\ref{eq:tbgamma}),
Eq.~(\ref{eq:Ggameq}) may easily be solved for $G_\mu^\gamma$
\begin{eqnarray}
&&G^\gamma_\mu(k_1',k_2';k_1,k_2;Q)=\int \frac{d^4p_1 d^4p_2 }{(2
\pi)^8} G(k_1',k_2';p_1+Q,p_2) j^{(1)}_\mu(Q^2) d_2^{-1}(p_2)
G(p_1,p_2;k_1,k_2)\nonumber\\ 
&& \quad + (1 \leftrightarrow 2) + \int
\frac{d^4p_1' d^4p_2' d^4p_1 d^4p_2}{(2 \pi)^{16}}
G(k_1',k_2';p_1',p_2') K^\gamma_\mu(p_1',p_2';p_1,p_2;Q)
G(p_1,p_2;k_1,k_2).
\label{eq:Ggameq2}
\end{eqnarray}
(Note that there is an overall delta function $\delta^{(4)} (k_1' +
k_2' - k_1 - k_2 - Q)$ on both sides of this equation.) Using the
identity (\ref{eq:G0gamma}) in Eq.~(\ref{eq:Ggameq2}) and the explicit
form of Eq.~(\ref{eq:GBSE}) we find
\begin{eqnarray}
&&Q^\mu G^\gamma_\mu(k_1',k_2';k_1,k_2;Q)=e_1 G(k_1'-Q,k_2';k_1,k_2) -
G(k_1',k_2';k_1+Q,k_2) e_1 \nonumber\\
&& + \int \frac{d^4p_1' d^4p_2' d^4p_1 d^4p_2}{(2 \pi)^{16}}
G(k_1',k_2';p_1',p_2')
\left[K(p_1',p_2';p_1+Q,p_2) e_1 \right.\nonumber\\
&& \qquad \qquad \qquad \qquad \qquad \qquad 
\left. - e_1 K(p_1'-Q,p_2';p_1,p_2)\right]
G(p_1,p_2;k_1,k_2) + (1 \leftrightarrow 2)\nonumber\\
&& + \int \frac{d^4p_1' d^4p_2' d^4p_1 d^4 p_2}{(2 \pi)^{16}}
G(k_1',k_2';p_1',p_2')
Q^\mu K_\mu^\gamma(p_1',p_2';p_1,p_2;Q) G(p_1,p_2;k_1,k_2).\nonumber\\
\end{eqnarray}
This reduces to a WTI for $G_\mu^\gamma$, 
\begin{equation}
Q^\mu G^\gamma_\mu(k_1',k_2';k_1,k_2;Q)=e_1 G(k_1'-Q,k_2';k_1,k_2) -
G(k_1',k_2';k_1+Q,k_2) e_1 + (1 \leftrightarrow 2),
\end{equation}
provided that
\begin{equation}
e_1 K(p_1'-Q,p_2';p_1,p_2) - K(p_1',p_2';p_1+Q,p_2) e_1 + (1
\leftrightarrow 2)=Q^\mu K^\gamma_\mu (p_1',p_2';p_1,p_2;Q),
\label{eq:cancellation}
\end{equation}
which is the WTI for the interaction current.  (Similar identities are
used in the construction of a gauge invariant electromagnetic matrix
element for the Gross---or spectator---formalism
in Refs.~\cite{GR87,KB97A}.) The result (\ref{eq:cancellation}) is
completely general, and will always hold if the two-body current
$K^\gamma_\mu$ is constructed in a gauge invariant way.  In
Appendix~\ref{ap-gimec} we show how Eq.~(\ref{eq:cancellation}) is
achieved in the special case where the photon couples only to particle
one, i.e. the exchanged particles and particle two carry no charge. By
contrast, if the particles are, for instance, nucleons carrying
isospin which exchange isovector mesons, the analysis is more
complicated, partly because the isospin operators in $K$ and $G$ do
not commute with those in the charges $e_1$ and $e_2$. However, as
alluded to above, the additional terms generated because of this
non-commutativity are canceled by terms arising from the coupling of
the photon to isovector mesons. (The interested reader may consult
Ref.~\cite{GR87} for details on this problem.)

Now, in the case of a bound state we have the following decomposition
of the Green's function
\begin{eqnarray}
&&G(k_1',k_2';k_1,k_2)=\nonumber\\
&& \quad \delta^{(4)}(k_1'+k_2'-k_1-k_2) \left[
G_0(k_1',k_2') \frac{\Gamma(k_1',k_2')
\bar{\Gamma}(k_1,k_2)} {P^2 - M^2} G_0(k_1,k_2) + R(k_1',k_2';k_1,k_2)
\right],
\label{eq:decomp}
\end{eqnarray}
where $P=k_1+k_2=k_1'+k_2'$ is the conserved total four-momentum of
the two-body system, $M$ is the two-body bound-state mass, and the
piece $R$ is regular at $P^2=M^2$. We then insert Eq.~(\ref{eq:decomp})
into (\ref{eq:Ggameq2}) and
extract the double-pole
part of the resulting expression, in order to obtain the amplitude for
interaction of the bound-state with a photon of momentum $Q$~\cite{Ma55}. 
Expressing the result in terms of total and relative four-momenta yields:
\begin{eqnarray}
&&{\cal A}_\mu(P,Q)=\int \frac{d^4p}{(2 \pi)^4} 
\bar{\Gamma}(p+Q/2;P+Q) {G_0^\gamma}_\mu (p;P,Q) 
\Gamma(p;P).\nonumber\\
&& \qquad + \int \frac{d^4p' \, d^4p}{(2 \pi)^8} 
\bar{\Gamma}(p';P+Q) G_0(p';P+Q) K_\mu^\gamma(p',P+Q;p,P;Q)
G_0(p;P) \Gamma(p;P).
\label{eq:gi4damp}
\end{eqnarray}
From Eq.~(\ref{eq:G0gamma}), Eq.~(\ref{eq:cancellation}), and the
bound-state BSE, Eq.~(\ref{eq:BSEbd}), it is straightforward to show that
\begin{equation}
Q^\mu {\cal A}_\mu(P,Q)=0,
\end{equation}
as required by gauge invariance.

In the context of the formalism of Section~\ref{sec-Section2} the
question is how to maintain this gauge invariance when the reduction
to three dimensions is made. In particular, in
Section~\ref{sec-Section2} we showed how $\Gamma_1$, which obeys a
three-dimensional integral equation, could be used to find
$\Gamma$. Inserting the expression (\ref{eq:Gamma}) into 
Eq.~(\ref{eq:gi4damp}) and using the definition of $\cal G$ we find
\begin{eqnarray}
&& {\cal A}_\mu(P,Q)=\int \frac{d^4k' d^4p' d^4p \, d^4k}{(2 \pi)^{16}} 
\bar{\Gamma}_1(k'_{\perp D};P+Q) 
{\cal G}(k',p';P+Q) \nonumber\\ 
&& \qquad \qquad \left[j_\mu^{(1)}(Q^2)
d_2^{-1}(\nu_2 P -p) \delta^{(4)}(p'-p-Q/2) + j_\mu^{(2)}(Q^2)
d_1^{-1}(\nu_1 P -p) \delta^{(4)}(p'-p+Q/2) \right.\nonumber\\ &&
\qquad \qquad \qquad \qquad \qquad \qquad \left.+ K^\gamma_\mu (P+Q,p';P,p;Q) 
\right] {\cal G}(p,k;P) \Gamma_1(k_{\perp D};P).
\label{eq:3damplfull}
\end{eqnarray}
An important part of Eq.~(\ref{eq:3damplfull}) is that we use the same
$D$ in the initial and final states.  From the discussion of
Section~\ref{sec-Section2}, it might appear that the natural choices
are $D = P$ in the initial state, and $D = P+Q$ in the final state.
However, if these choices are made then the formula
(\ref{eq:3damplfull}) is actually slightly misleading, because the two
vertex functions so defined depend on pieces of the four-vectors $k$
and $k'$ which reside in different subspaces. Thus, for instance, the
two ${\cal G}$'s appearing in Eq.~(\ref{eq:3damplfull}) are actually
different objects. This immediately leads to significant
complications. In particular it makes it very difficult to define a
three-dimensional gauge invariant approximation to the amplitude
(\ref{eq:gi4damp}).

In this current matrix element, the choice
\begin{equation}
D=(P + P')/2,
\label{eq:Dave}
\end{equation}
may be made in both the initial and final state. With this choice the
vertex functions $\Gamma_1(k_{\perp D}, P+Q)$ and $\Gamma_1(k_{\perp
D}, P)$ cannot both be obtained from a kinematical boost of the rest
frame wave functions with $D=(1,0,0,0)$.  However, in spite of the
resultant variation in the bound-state mass discussed in
Section~\ref{sec-Section2}, the matrix element (\ref{eq:3damplfull})
{\it is} covariant provided that {\it both} the vectors $P$ and $P'$
are boosted using the same Lorentz transformation ${\cal L}$. This is
easily shown by expanding Eq.~(\ref{eq:3damplfull}) in powers of
$K-K_1$ and then using Eq.~(\ref{eq:covG1Pt}), the definition of
$K_1$, and the known covariance properties of $K$ and $G_0$.

The choice (\ref{eq:Dave}) is a useful one because it allows the
straightforward definition of a three-dimensional matrix element for
the interaction of the bound state with an electromagnetic probe which
is not only covariant, but is also gauge invariant.  To show this,
first note that in order to solve the integral equation
(\ref{eq:3DIE}), we in fact choose $K_1$ so as to impose
\begin{equation}
\langle {\cal G} \rangle=\langle G_0 \rangle.
\label{eq:calGeq}
\end{equation}
If this condition is imposed to all orders in the coupling constant
then the result is an infinite series for $K_1$.  If we impose the
condition (\ref{eq:calGeq}) order-by-order in the expansion in
$K-K_1$, the condition defines $K_1$ to the same order.  Truncation of
the kernel is necessary for a practical analysis and it becomes
crucial to inquire if a corresponding approximation for the matrix
element (\ref{eq:3damplfull}) exists that maintains the gauge
invariance of the theory.  It turns out that the
current matrix element (\ref{eq:3damplfull}) is gauge invariant if
${\cal G} (J_\mu + K^\gamma_\mu) {\cal G}$ on the right-hand side of
Eq.~(\ref{eq:3damplfull}) is expanded to a given order in the coupling
constant and the kernel $K_1$ used to define $\Gamma_1$ is obtained
from Eq.~(\ref{eq:calGeq}) by truncation at
the same order in the coupling constant.

To see this we first abbreviate Eq.~(\ref{eq:3damplfull}) as:
\begin{equation}
{\cal A}_\mu(P,Q)=\bar{\Gamma}_1(P+Q) {\cal G}(P+Q) (J_\mu + K^\gamma_\mu)
{\cal G}(P) \Gamma_1(P),
\label{eq:Amu}
\end{equation}
where all integrations over relative momenta are now suppressed. We
split this into two pieces, one due to the one-body current $J_\mu$,
and one due to the two-body current $K^\gamma_\mu$. Suppose that $K_1$
has been truncated at lowest order, i.e. $K_1=K_1^{(2)}$.  Then, in
the $J_\mu$ piece, we expand the $\cal G$s and retain terms up to
first order in $K-K_1$. To this we add a piece from the two-body
current, in which we stop this expansion of $\cal G$ at zeroth order
in $K-K_1$, i.e., write ${\cal G}=G_0$.  Thus, we define a first-order
approximation to ${\cal A}_\mu$, ${\cal A}_\mu^{(1)}$, by
\begin{eqnarray}
{\cal A}^{(1)}_\mu&=&\bar{\Gamma}_1(P+Q) \langle G^\gamma_{0 \mu}
\rangle \Gamma_1(P) + \bar{\Gamma}_1(P+Q) \langle G_0(P+Q) (K(P+Q)-K_1(P+Q))
G^\gamma_{0 \mu} \rangle \Gamma_1(P)
\nonumber\\
&+& \bar{\Gamma}_1(P+Q) \langle G^{\gamma}_{0 \mu} (K(P)-K_1(P)) G_0(P) \rangle
 \Gamma_1(P)
\nonumber\\ 
&+& \bar{\Gamma}_1(P+Q) \langle G_0(P+Q) K^\gamma_\mu G_0(P) \rangle
\Gamma_1(P).
\label{eq:A1mu}
\end{eqnarray}
It follows from Eq.~(\ref{eq:covG1Pt}) that ${\cal A}_\mu^{(1)}$ is a
covariant matrix element.  To show ${\cal A}^{(1)}_\mu$ is also gauge
invariant we contract with the four-vector $Q$ and use the identities
(\ref{eq:G0gamma}) and (\ref{eq:cancellation}) This leads to
\begin{eqnarray}
Q^\mu {\cal A}^{(1)}_\mu&=&\bar{\Gamma}_1(P+Q) e_1 \langle G_0(P) \rangle
\Gamma_1(P)
- \bar{\Gamma}_1(P+Q) \langle G_0(P+Q) \rangle e_1 \Gamma_1(P)\nonumber\\
&+& \bar{\Gamma}_1(P+Q) \langle G_0(P+Q) (K(P+Q)-K_1(P+Q)) e_1 G_0(P)
\rangle
\Gamma_1(P)\nonumber\\ 
&-& \bar{\Gamma}_1(P+Q) \langle G_0(P+Q)
(K(P+Q)-K_1(P+Q)) G_0(P+Q) \rangle e_1 \Gamma_1(P)\nonumber\\ 
&+&\bar{\Gamma}_1(P+Q) e_1 \langle G_0(P) (K(P)-K_1(P)) G_0(P) \rangle
\Gamma_1(P)\nonumber\\ 
&-& \bar{\Gamma}_1(P+Q) \langle G_0(P+Q) e_1 (K(P)-K_1(P))
G_0(P) \rangle \Gamma_1(P)\nonumber\\
&+& \bar{\Gamma}_1(P+Q) \langle G_0(P+Q) (e_1 K(P) - K(P+Q) e_1) G_0(P)\rangle 
\Gamma_1(P) + (1 \leftrightarrow 2).
\end{eqnarray}
The bound-state equation (\ref{eq:K1eqn}) can now be used to show
that the terms in the first two lines cancel the terms in the last two lines,
and so:
\begin{eqnarray}
Q^\mu {\cal A}^{(1)}_\mu=
-\bar{\Gamma}_1(P+Q) \langle G_0(P+Q) (K(P+Q)-K_1(P+Q)) G_0(P+Q) \rangle e_1
\Gamma_1(P)\nonumber\\
+ \bar{\Gamma}_1(P+Q) e_1 \langle G_0(P) (K(P)-K_1(P)) G_0(P) \rangle
\Gamma_1(P)+ (1 \leftrightarrow 2).
\end{eqnarray}
But, at second-order in the coupling constant, Eq.~(\ref{eq:calGeq})
leads to Eq.~(\ref{eq:K1Klein}) for $K_1$. It immediately follows that
if $K_1$ has been defined in this way, the corresponding amplitude for
electromagnetic interactions of the bound state, as defined by
Eq.~(\ref{eq:A1mu}), obeys
\begin{equation}
Q^\mu {\cal A}^{(1)}_\mu=0.
\end{equation}

It is straightforward to check that the same result holds if
Eq.~(\ref{eq:K1eqn}) is truncated at next order, while the one-body
current pieces of Eq.~(\ref{eq:3damplfull}) are expanded to second
order in $K-K_1$ and the two-body current pieces are expanded to first
order in $K-K_1$. That is, if the interaction $K_1^{(4)}$ defined by
(\ref{eq:K14}) is used, then defining a vertex function $A_\mu^{(2)}$
by expanding Eq.~(\ref{eq:Amu}) to fourth order in the coupling
constant leads to $Q^\mu {\cal A}^{(2)}_\mu=0$ also. Thus, we conclude
that truncating the kernel $K_1$ defined by Eq.~(\ref{eq:calGeq}) and
the electromagnetic vertex defined by Eq.~(\ref{eq:Amu}) at
consistent order in the coupling constant yields a gauge invariant,
covariant, electromagnetic matrix element.

\begin{figure}[h]
\centerline{\BoxedEPSF{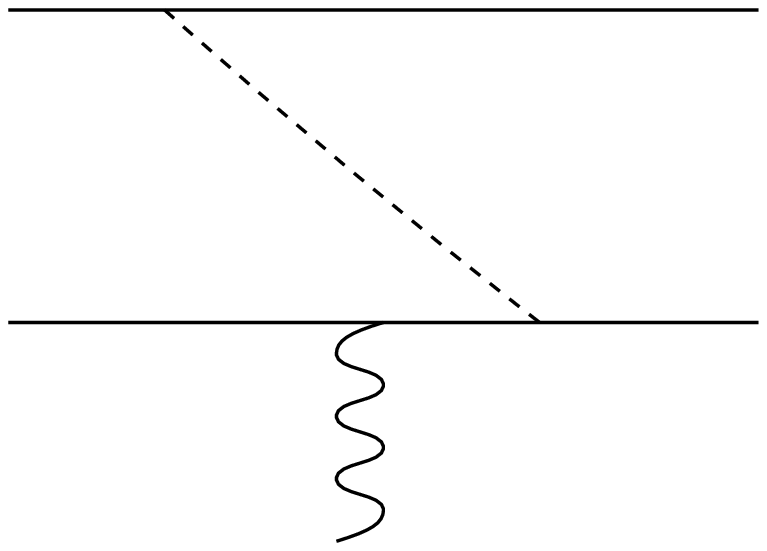 scaled 700}}
\caption{One example of a graph which is required in our formalism
in order to maintain gauge invariance.}
\label{fig-fig1}  
\end{figure}

It should be pointed out that ${\cal A}_\mu^{(1)}$ includes
contributions from diagrams where the photon couples to particles one
and two while exchanged quanta are ``in-flight''. These contributions
are of two kinds.  Firstly, if the four-dimensional kernel $K$ is
dependent on the total momentum then gauge invariance requires the
presence of terms representing the coupling of the photon to internal
lines in $K$.  Secondly, even if the kernel does not depend on the
total four-momentum, e.g. it is a OBE kernel, terms arise in the
three-dimensional formalism where the photon couples to particles one
and two while an exchanged particle is ``in-flight''. These must be
included if our approach is to contain a WTI. (See Fig.~\ref{fig-fig1}
for a diagrammatic interpretation of one such term.)  Many years ago,
in the course of work on deuteron photo-disintegration, Pearlstein and
Klein~\cite{PK60} showed that contributions such as these had to be
included if Siegert's theorem (which is a consequence of gauge
invariance) was to hold in the Klein approach.  Our systematic
derivation shows the general form of the additional pieces of the
current operator which are required if gauge invariance is to be
maintained.

Gross and Riska~\cite{GR87} and Kvinikhidze and
Blankleider~\cite{KB97A} have discussed the construction of gauge
invariant amplitudes for the interaction of a photon with the
two-body system within the Gross (or spectator) quasipotential
approach~\cite{Gr82,vO95A}.  It would be interesting to compare the
systematic approach to constructing electromagnetic matrix elements
developed here with the discussions of Refs.~\cite{GR87,KB97A}.

\section {Conclusion}

In this paper, a systematic formalism is presented for the reduction
of the Bethe-Salpeter equation to a three-dimensional form in which
components of the relative momentum parallel to a direction $D$ are
integrated out. By choosing $D=P$, where $P$ is the total four-momentum
of the two-body system, a covariant formalism is obtained for a bound state.

This part of the development follows closely the analysis of Ref.~\cite{PW96},
where the corresponding formalism based on the choice $D=(1,0,0,0)$
was constructed. However, here we simplify the analysis by omitting the
treatment of crossed graphs in order to focus on covariance and
gauge invariance.

Consideration of the matrix elements of a conserved current shows that
the gauge invariance of the underlying four-dimensional theory is preserved
in the three-dimensional reduction. Moreover, it is possible to truncate both 
the infinite series expansion for the interaction kernel used to define 
the bound states and the expansion of the currents at the same order 
in the coupling constant, and maintain this gauge invariance.

The three-dimensional reduction for current matrix elements is based upon the 
use of the same direction $D$ in initial and final states. Here, the choice 
$D=\frac{1}{2}(P+P')$ is suggested, where $P$ ($P'$) is the initial (final)
total four-momentum of the two-body system. Other possible choices,
e.g. $D=P$ or $D=P'$, could be used just as well, and may be preferable
in some circumstances. The current matrix elements are
covariant for any choice of $D$ that is a linear
combination of the two independent four-vectors $P$ and $P'$.

When truncations of the interaction kernel are employed in order to
implement a practical calculation the results depend on the
choice of $D$. The dependence on $D$ in a truncated kernel, $K_1$, can
be reduced systematically by keeping higher-order
terms. Alternatively, the invariance of the mass of initial and final
bound states can be enforced by a rescaling of the coupling constants
in $K_1$.

There are two important advantages of the reduction formalism based on
integrating out momentum components presented here. Firstly, it is
systematic. Secondly, as discussed in Ref.~\cite{PW96}, it avoids
unphysical singularities. Such singularities generally arise in
systematic quasipotential reductions to three dimensions. Typically
such reductions involve splitting the propagator into a part involving
a delta function and a remainder. Corrections to the theory have
unphysical singularities, such as particle production thresholds at
energies below the physical values. Moreover, current matrix elements
obtained using quasipotential reductions generally involve similar
unphysical singularities which must be either disregarded or
suppressed, with uncertain consequences. Such difficulties have
motivated the present approach. This systematic formalism in which
gauge invariance can be maintained by consistent approximations in the
current operator and interaction kernel should prove to be a useful tool
in the calculation of relativistic bound state properties.

\acknowledgements{We are grateful to the U.~S. Department of Energy
for its support under grant no. DE-FG02-93ER-40762. We also both thank
Professor Tjon for a number of useful and stimulating conversations
on this topic, and on other topics in relativistic few-body physics.
D.~R.~P. thanks Iraj Afnan for useful correspondence on this paper.}

\appendix

\section{Proof of Ward-Takahashi identity for $K_\mu^\gamma$ in a special case}

\label{ap-gimec}

In what follows we assume that the photon couples only to particle
one. If the interaction between particles one and two is mediated by
quanta which carry charge, then the discussion given here must be
modified to incorporate terms where the photons couple to the
exchanged quanta. For instance, if particles one and two are nucleons
carrying isospin which exchange isovector mesons, the analysis is more
complicated, due to the non-commutativity of isospin operators in $K$
and $G$ with those in the charges $e_1$ and $e_2$. However, the
additional terms generated via this complication are canceled by
terms arising from the direct coupling of the photon to the isovector
mesons~\cite{GR87}.

Consider the $n$th order piece of the kernel $K$. If $e_2=0$ and the
exchanged particles do not couple to the photon we need only concern
ourselves with the particle one pieces of the three terms in
Eq.~(\ref{eq:cancellation}), since the particle two and meson
propagators may be arranged so as to take exactly the same form in all
three pieces. Suppose then that the momenta on internal lines of
particle one are labeled $p_1^{(1)}, p_1^{(2)}, \ldots ,p_1^{(j)}$ to
the right of the photon insertion, and $p_1^{(j)}+Q, p_1^{(j+1)}+Q,
\ldots ,p_1^{(n-1)}+Q$ to the left of the photon insertion. Meanwhile,
the vertices governing the interaction of the exchanged quanta with
particle one are labeled $V_1^{(1)}, \ldots, V_1^{(n)}$. With these
conventions we may write the particle one pieces of the three terms in
Eq.~(\ref{eq:cancellation}) as:
\begin{eqnarray}
K(p_1',p_2';p_1+Q,p_2)&:& V_1^{(n)} \prod_{i=1}^{n-1} d_1 (p_1^{(i)} + Q)
V_1^{(i)}\\ K(p_1'-Q,p_2';p_1,p_2) &:& V_1^{(n)} \prod_{i=1}^{n-1} d_1
(p_1^{(i)}) V_1^{(i)}\\ 
K_\mu^\gamma (p_1',p_2';p_1,p_2;Q) &:&
\sum_{j=1}^{n-1} V_1^{(n)} \left[ \prod_{i=j+1}^{n-1} d_1
(p_1^{(i)} + Q) V_1^{(i)} \right] g_{0 \mu}^{(1)} (p_1^{(j)},Q)
V_1^{(j)} \left[ \prod_{i=1}^{j-1} d_1 (p_1^{(i)}) V_1^{(i)} \right],
\end{eqnarray}
where the product symbols are defined to give one if the upper index
of the product lies below the lower index. Note that the changes in
particle one's external momenta guarantee that the particle two and
meson pieces of the diagram will be identical in all three cases.

Now calculating $Q^\mu K_\mu^\gamma$ using the identity (\ref{eq:obWTI}) 
we obtain, for the particle one pieces of this object:
\begin{eqnarray}
&& \sum_{j=1}^{n-1} V_1^{(n)} \left[ \prod_{i=j+1}^{n-1} d_1
(p_1^{(i)} + Q) V_1^{(i)} \right] e_1 d_1 (p_1^{(j)}) V_1^{(j)} \left[
\prod_{i=1}^{j-1} d_1 (p_1^{(i)}) V_1^{(i)} \right]\nonumber\\
&& \qquad \qquad - \sum_{j=1}^{n-1} V_1^{(n)} \left[ \prod_{i=j+1}^{n-1} d_1
(p_1^{(i)} + Q) V_1^{(i)} \right] d_1 (p_1^{(j)} + Q) e_1
V_1^{(j)} \left[ \prod_{i=1}^{j} d_1
(p_1^{(i)}) V_1^{(i)} \right].
\end{eqnarray}
This is a telescoping series, as the $j$th term from the first piece
cancels the $j+1$th term from the second piece.  The only structures
which survive this cancelation give precisely the particle one piece
of 
\begin{equation}
e_1 K(p_1'-Q,p_2';p_1,p_2) - K(p_1',p_2';p_1+Q,p_2) e_1.
\end{equation}
Note that the series can only be summed in this way if the
vertices commute with the charge operator $e_1$. However, if the
particles are distinguishable and $[V_1^{(j)},e_1]=0$ then this
argument suffices to prove Eq.~(\ref{eq:cancellation}).

\end{document}